\documentclass[prl,showpacs,twocolumn,  preprintnumbers, amsmath, amssymb]{revtex4}
\usepackage{graphicx}% Include figure files
\usepackage{dcolumn}% Align table columns on decimal point
\usepackage{bm}% bold math

\begin{document}
\title{In Search of Microscopic Evidence for Molecular Level Negative Thermal Expansion in Fullerenes}

\author{S. Brown$^1$}
\email{brown@ion.chem.utk.edu}
\author{J. Cao$^1$}
\author{J. L. Musfeldt$^1$}
\author{N. Dragoe$^2$}
\author{F. Cimpoesu$^3$}

\author{S. Ito$^4$}
\author{H. Takagi$^4$}
\author{R. J. Cross$^5$}

\affiliation{$^1$Department of Chemistry, University of Tennessee,
Knoxville, Tennessee 37996-1600}

\affiliation{$^2$ICMMO - LEMHE, Universit\'{e} Paris XI, Bat 410,
UMR 8647-CNRS, Orsay 91405, France}

\affiliation{$^3$Institute of Physical Chemistry, Splaiul
Independentei 202, Bucharest 77208, Romania }

\affiliation{$^4$Department of Advanced Materials Science,
University of Tokyo, Kashiwa-no-ha 5-1-5 Kibanto 403, Kasiwa
277-8561, Japan}

\affiliation{$^5$Department of Chemistry, Yale University, PO Box
208107, New Haven, Connecticut 06520-8107}

\date{\today}

\begin{abstract}
We report the high-resolution far infrared vibrational properties of
C$_{60}$ and endohedral Kr@C$_{60}$ fullerene as a function of
temperature. Anomalous softening of the F$_{1u}(1)$ mode (526
cm$^{-1}$) is observed throughout the temperature range of
investigation (300 - 10 K) suggesting that the fullerene cage may
expand at low temperature in these molecular solids. To test this
idea, we combine these results with temperature and pressure
dependent Raman, infrared, and Kr extended x-ray absorption fine
structure data from the literature to provide a comprehensive view
of local cage size effects. The results are consistent with a recent
molecular dynamics study [Kwon \emph{et al.}, Phys. Rev. Lett.
\textbf{92}, 15901 (2004)] that predicts negative thermal
expansion in carbon fullerenes.\\

\end{abstract}


 \narrowtext

\maketitle

\clearpage

\section{Introduction}

The interplay between thermal, mechanical, and vibrational
properties that gives rise to negative thermal expansion (NTE) in
complex materials is an area of sustained scientific interest.
ZrW$_2$O$_8$ is a prime example, where bulk NTE is present over a
wide temperature range and is attributed to unusual low-frequency
dynamics \cite{Hancock2004, Mary1996, Ernst1998, Ravindran2000}.
 Various layered and molecular solids exhibit
similar behavior although the mechanisms driving NTE can be
different. For instance, graphite and related layered solids display
anisotropic negative thermal expansion even well above room
temperature \cite{Barrera2005,Nelson1945}.
 In graphite, NTE is primarily due to the intra-layer
tension effect, whereas in molecular solids such as
Sm$_{2.72}$C$_{60}$, NTE originates from the mixed electronic
configuration of Sm \cite{Arvanitidis2003}. Based on bonding pattern
similarities  and compressibility arguments, it has been anticipated
that other nanocarbon materials (including fullerenes, small cage
compounds, and nanotubes) may also exhibit unusual properties
\cite{Kwon2004}. Independent of the many scientific discoveries in
this area, the technological dream is to exploit negative and zero
expansion materials in composites and molecular electronics devices
that have specifically tailored thermal expansion coefficients.

A number of recent molecular dynamics simulations have focused on
the thermal properties of small cage fullerenes and carbon nanotubes
\cite{Kwon2004,Li2005,Jiang2004,Schelling2003}. Of particular
interest is the molecular dynamics investigation by Kwon \emph{et
al.}, which predicted
 NTE in C$_{60}$ up to moderate temperatures
 due to an unusual combination of structural
and vibrational entropic effects  \cite{Kwon2004}. The calculated
value of the maximum volumetric thermal expansion coefficient is -1$
\times$10$^{-5}$ K$^{-1}$ \cite{Kwon2004}, comparable to that in
ZrW$_2$O$_8$ \cite{Mary1996}. It is important to note that NTE in
C$_{60}$ is predicted to be a molecular-level rather than bulk
effect \cite{Kwon2004}. As a consequence, local or ``microscopic"
probes are required to investigate this prediction. At this time,
theoretical results for carbon nanotubes are more controversial
\cite{Li2005, Kwon2004, Jiang2004, Schelling2003}, demonstrating the
need for additional  work in this area.

At ambient temperature and pressure,  C$_{60}$ fullerene  has the
shape of a truncated icosahedron (point group I$_h$) with 12
pentagonal and 20 hexagonal faces. It displays both long (between a
hexagon and a pentagon, $l$= 1.45 ${\AA}$) and short (between
hexagons, $l$= 1.40 ${\AA}$) bonds \cite{Yannoni1991a}. Due to the
high point symmetry, isolated C$_{60}$ molecules display 46 distinct
vibrational modes. Of this set, 4F$_{1u}$ are infrared active, and
2A$_g$ + 8 H$_g$ are Raman active. These modes can be differentiated
by the percentage of radial and tangential displacement of molecule
in a particular vibrational mode \cite{Stanton1988}. For example,
 F$_{1u}$(1) and F$_{1u}$(2) modes are 93.5 and 66.6\%  radial while
F$_{1u}$(3) and F$_{1u}$(4) are 48.7 and 99.5\% tangential in
character, respectively \cite{Stanton1988, F1u1}. In the solid
state, C$_{60}$ molecules crystallize in face-centered cubic
arrangement with space group $Fm\bar{3}m$ (O$_h$$^5$)
\cite{Heiney1992}. The room temperature phase is orientationally
disordered due to rapid and continuous rotation of the C$_{60}$
icosahedra \cite{Yannoni1991b}. C$_{60}$ undergoes a first-order
structural transition to a simple cubic structure
($Pa\bar{3}$/$T_h$$^6$) at 260 K and to an orientationally ordered
phase at around 90 K {\cite{Sachidanandam1991,David1991,Heiney1991}.
 Bulk structural studies show that lattice parameter trends  are standard
over the full temperature regime (Table~\ref{tab_temp})
\cite{260Knote}.

Incorporation of a noble gas atom such as Kr inside the cage leads
to a stable endohedral fullerene (Kr@C$_{60}$) with an average Kr-C
separation of 3.540(3) {\AA} \cite{Lee2002, Ito2004}. Comparison of
this distance with the sum of the van der Waals radii for carbon and
krypton (3.72 \AA) demonstrates that the Kr atom is tightly confined
within the surrounding cage. Recent synchrotron x-ray diffraction
work on the similar Ar@C$_{60}$ indicates that the unit cell
parameter  is slightly larger  than that of C$_{60}$ at 300 K and
slightly smaller at low temperature \cite{Takeda2005}.

 \begin{table}[t]
   \caption{\label{tab_temp}}%\vspace{0.2cm}
   Structural studies of crystalline C$_{60}$ fullerene. \\
   \begin{ruledtabular}

   \begin{tabular}{llrllc}
   T (K) & a ({\AA}) & Ref. & P (GPa)& a ({\AA}) & Ref. \\
   \hline
   \\
 %  300 & 14.159 &\cite{David1993} & &  \\
   270 & 14.1543(2) &\cite{David1993}& 0.35(5) &14.067(2)&\cite{Jephcoat1994}\\
   170 & 14.0708(1)  &\cite{David1993}  &0.41(5) &14.031(5)&\cite{Jephcoat1994}\\
   110 & 14.052(5)  &\cite{Liu1991}    &0.55(5) &13.978&\cite{Jephcoat1994}\\
   11 & 14.04(1)  & \cite{Heiney1991}   &2.4(1) &13.730&\cite{Jephcoat1994}\\
   5 &14.0408(1)    &\cite{David1991}   &2.5(1) &13.633&\cite{Jephcoat1994}\\

   \end{tabular}
   \end{ruledtabular}
   \end{table}

It is clearly of great interest to compare the theoretical
predictions for molecular level NTE in small cage fullerenes with
 experimental observations.  In this article, we report a
 high-resolution far infrared study of C$_{60}$ and Kr@C$_{60}$ as a
 function of temperature with particular focus on the F$_{1u}(1)$ and
 F$_{1u}(2)$ vibrational modes. Unusual softening of the F$_{1u}(1)$ mode is observed
with decreasing temperature in both compounds, consistent with
predictions for an expanded low temperature fullerene ball. To
further test this idea, we combine our results with
 previous structural and vibrational studies (including variable temperature and pressure Raman, infrared, and Kr
extended x-ray absorption fine structure data
\cite{Loosdrecht1992,Matus1993,Hamanaka1995,Horoyski1995,Meletov1995a,
Chandrabhas1992,Snoke1992,Aoki1991,Martin1994,Meletov1995b,Huang1991,Ito2004})
to evaluate  mode Gr\"uneissen parameters  and elucidate the thermal
expansion coefficient of C$_{60}$. Many of the mode Gr\"uneissen
parameters are negative, and we estimate that $\alpha$$_{vib}$  is
on the order of -10$^{-5}$
 K$^{-1}$. We also discuss the subtle guest-host interaction in
Kr@C$_{60}$.

\section{Methods}

Endohedral fullerenes were synthesized using the high-pressure,
high-temperature technique described
 earlier \cite{Ito2004}. The resulting product contained both  C$_{60}$ powder
 and $\sim$1\% Kr@C$_{60}$. Multiple separations were carried
 out using a Jasco Gulliver 1500 high-performance liquid
 chromatography system, equipped with recycling,
 automatic injection, and a diode array detector (300 - 900 nm).
Using a detailed procedure involving recycling and sequential
chromatography \cite{Ito2004}, about 1 mg of 99\% pure Kr@C$_{60}$
was obtained.  The pristine C$_{60}$ used in this work was obtained
from Aldrich (99.9\% purity).

Variable temperature far infrared transmittance measurements were
performed using a Bruker 113V Fourier transform infrared
spectrometer, covering the frequency range from 22 - 650 cm$^{-1}$
with both 0.1 and 0.5 cm$^{-1}$ resolution.  A helium-cooled
bolometer detector was employed  for added sensitivity. The
measurements were done using isotropic, pressed pellets ($\sim$1\%
 by weight) with paraffin as the matrix material. Low temperature
 spectroscopies
were carried out with a continuous-flow helium cryostat and
temperature controller. Standard peak fitting techniques were
employed, as appropriate.

\section{Results and Discussion}

\begin{figure}[t]
  \includegraphics[width = 3.0 in]{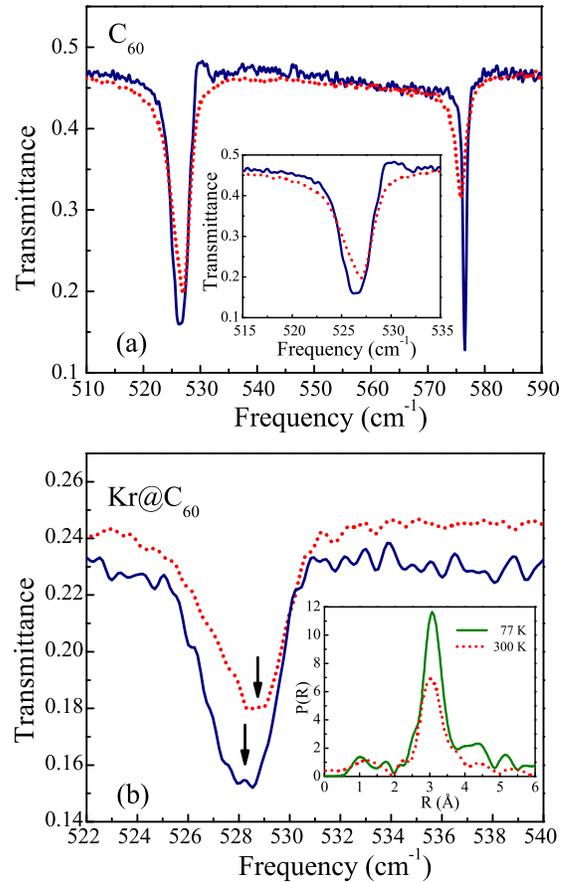}% Here is how to import EPS art
  \caption{\label{fig_softening}(Color online)
   (a) High resolution far infrared
   transmittance spectra of C$_{60}$. Note that the F$_{1u}$(1) mode splits and
   softens at low temperature,
   whereas the
   F$_{1u}$(2) mode hardens.  The inset shows a
   close-up view of F$_{1u}$(1) mode softening. (b) Close-up view of the
   low temperature F$_{1u}$(1) mode softening in Kr@C$_{60}$. The inset
   displays previous
    Kr-EXAFS results \cite{Ito2004}.
     In all panels, the
   red (dotted) lines represent 300 K results, whereas the blue (solid)
   lines correspond to 4 K data. In the inset, the green (solid) line
    corresponds to 77 K data.}
  \end{figure}

Figure ~\ref{fig_softening} shows the far infrared transmittance
spectra of fullerene C$_{60}$ and Kr@C$_{60}$ at 10  and 300 K. For
C$_{60}$, the radial F$_{1u}$(1) and F$_{1u}$(2) modes are observed
at 526 and 576 cm$^{-1}$, consistent with previous work
\cite{Homes1994}. The most striking result is the temperature
dependence of the F$_{1u}$(1) mode, which softens by $\sim$0.5
cm$^{-1}$
 with decreasing temperature. This effect occurs gradually throughout
 the entire temperature range.
 Note that the fine structure in F$_{1u}$(1) is a result of crystal field
symmetry breaking below the 260 K transition and was investigated in
detail by Homes \emph{et al.} \cite{Homes1994}.  At low temperature,
F$_{1u}$ splits as F$_{1u}$ $\rightarrow$ A$_u$ + E$_u$ + 3F$_u$, of
which only the F$_u$ modes are infrared active
\cite{spacegroup,DongNote}. The unusual softening of F$_{1u}$(1) was
not, however,  discussed in earlier measurements. A similar
softening effect appears in endohedral Kr@C$_{60}$, as shown in Fig.
~\ref{fig_softening}(b). Taken together, these temperature dependent
mode softening trends support a more relaxed ball at low
temperature. In contrast to the observed softening  in F$_{1u}$(1),
the other radial mode, F$_{1u}$(2), exhibits normal behavior, that
is, hardening with decreasing temperature. The mode sharpens and
grows in strength as the sample is cooled. The absence of crystal
field-induced fine structure in F$_{1u}$(2) has been previously
attributed to the degree of overlap between 6:6 bond and the
adjacent pentagon \cite{Narasimhan1992}.

Figure~\ref{fig_trend} displays the peak position of the F$_{1u}$(1)
mode in pristine C$_{60}$ and Kr@C$_{60}$ as a function of
temperature. Two major trends are observed. First, the F$_{1u}$(1)
mode softens throughout the measured temperature range for both
 C$_{60}$ and Kr@C$_{60}$. Second, the F$_{1u}$(1) mode hardens
  by $\sim$2 cm$^{-1}$ upon endohedral incorporation \cite{Note2}. A hardening trend was also observed for  the F$_{1u}$(1) mode in a
previous study \cite{Yamamoto1999}. The F$_{1u}$(2) mode frequency
in Kr@C$_{60}$ (577 cm$^{-1}$) is also $\sim$1 cm$^{-1}$ higher than
that in the pristine material. Noble gas-containing endohedrals are
traditionally considered to be van der Waals complexes due to the
inert nature of noble gas guest atom. In such a picture, only modest
spectral changes (such as the appearance of a rattling mode) are
anticipated upon endohedral incorporation. The observed radial mode
hardening (Fig. \ref{fig_trend}) and tangential mode softening
\cite{Yokoyama} in Kr@C$_{60}$ compared with pristine C$_{60}$
suggests a modest guest-host interaction and a slight modification
of the encapsulant/cage interaction potential.

Returning our attention to the temperature dependent  mode softening
problem, a literature survey shows that a number of Raman active
modes also display unexpected low temperature softening
\cite{Hamanaka1995, Horoyski1995, Loosdrecht1992, Matus1993}. Just
to cite a few examples, van Loosdrecht \emph{et al.}
\cite{Loosdrecht1992} observed softening of several different modes
(H$_g$(3), H$_g$(4), A$_g$(2) and H$_g$(8)) both above and below the
260 K transition temperature. (All modes harden at the transition
temperature.) Hamanaka \emph{et al.} showed that the H$_g$(7) mode
softens below 260 K, whereas the A$_g$(2) mode softens below 160 K.
The H$_g$(3) and H$_g$(5) modes remain almost unchanged with
decreasing temperature \cite{Hamanaka1995}. In contrast, Horoyski
\emph{et al.}
  reported almost no change in A$_g$(2) and H$_g$(4) in the lower temperature
range  \cite{Horoyski1995, Comment1}.  Dong \emph{et al.}  observe
higher order overtone and combination modes in C$_{60}$, although
their exact temperature dependence is not clear \cite{Dong1993}.
Taken together, we find that several (but not all) Raman
investigations support unusual low temperature vibrational mode
softening.

\begin{figure}
\includegraphics[width = 3.0 in]{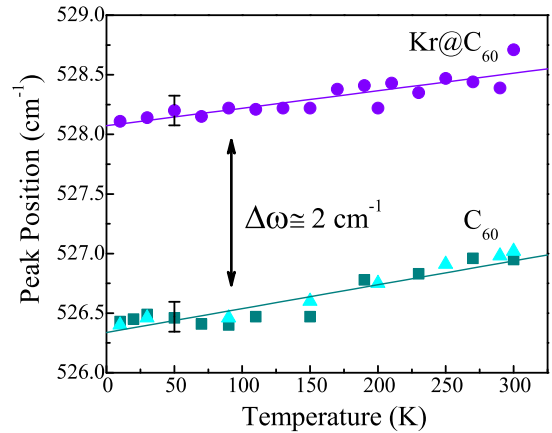}% Here is how to import EPS art
\caption{\label{fig_trend} (Color online) Temperature dependence of
the center peak position of the F$_{1u}$(1) mode in C$_{60}$ and
Kr@C$_{60}$. The blue-shift of this mode in the endohedral material
suggests modest guest-host interactions. Two independent runs are
shown for C$_{60}$. Error bars are indicated.}

\end{figure}

It is of interest to understand the microscopic origin of the
unusual mode softening and its implications for the bulk properties.
 In general, mode softening can be correlated to two major
effects: volume and charge. The charge effect is well-known in
organic charge transfer salts \cite{Swietlik2004, Wang1994}.
 Typically, volume contributions dominate
 in high pressure experiments. High-pressure vibrational
studies, detailed below, also indicate extensive mode softening
\cite{Meletov1995a, Snoke1992, Chandrabhas1992, Martin1994,
Meletov1995b, Huang1991, Aoki1991}. Hence, we attribute both
pressure and temperature dependent mode softening  to cage volume
effects \cite{Comment4}.

Volume change can be quantified by the thermal expansion coefficient
as

\begin{equation}
\alpha = \frac{\gamma_{av} C_V}{V_mB} = \frac{\chi_T}{V_m }\sum_i
\gamma_i c_i.
\end{equation}

\noindent
 Here, B (= 1/$\chi_T$) is the
bulk modulus,  $\gamma_{av}$ (= $\sum_i \gamma_i c_i / \sum_i c_i$)
is the average of mode Gr\"{u}neisen parameters $\gamma_i$  weighted
by the mode specific heat $c_i$, $C_V$ is total molar specific heat
at constant volume, and $V_m$ is the molar volume \cite{Barron1999}.
The mode Gr\"{u}neisen parameters are dimensionless quantities that
relate the fractional change in volume to the fractional change in
frequency of a given mode. They are defined as

\begin{equation}
\gamma_i = -\left(\frac{\delta  \ln  \omega_i}{\delta \ln V }\right)
= \frac{1}{\omega_i\chi_T} \left(\frac{\delta \omega_i} { \delta
P}\right),
\end{equation}

\noindent
 where $\omega_i$ is the frequency of the $i^{th}$ mode. At high
 temperatures, the  $\gamma_{av}$ can be taken as the arithmetic
 mean of mode Gr\"{u}neisen parameters $\gamma_i$.
Note that thermal expansion will be positive or negative depending
upon whether positive or negative $\gamma_i$ predominate in the
weighted average \cite{Barron1999, Ernst1998, Ravindran2000}.

To evaluate these quantities, we examined the pressure dependent
vibrational properties data that are available in the literature.
Meletov \emph{et al.}
 measured the Raman spectrum of C$_{60}$ between 200 - 800 and
 1350 - 1700 cm$^{-1}$
 \cite{Meletov1995a,Comment5}.
 The results clearly show  three well-defined phases: below 0.4 GPa,
 between 0.4 and 2.4 GPa, and  above 2.4 GPa.
 These pressure-induced phase transitions are attributed to orientational
 ordering from face-centered cubic to a simple cubic structure and
 finally to a rotation-free orientationally-ordered simple cubic phase in
 the high pressure regime \cite{Meletov1995a}. Comparison of lattice parameter trends
 with temperature and pressure  (Table~\ref{tab_temp}) demonstrates the rationale
 for concentrating
 our analysis on the pressure range below 0.4 GPa, although
 to obtain the fullest
 possible mode contributions, we also employ higher pressure results in our secondary
 analysis, below.
Returning to the data of Meletov \emph{et al.} \cite{Meletov1995a},
almost all the modes soften up to 0.4 GPa, whereas H$_g$(3),
H$_g$(4) soften through the higher pressure range as well.
 Similar softening has been observed for the A$_g$(2) mode at 0.35 GPa \cite{Chandrabhas1992}.
 Further, Snoke \emph{et al.} showed that the F$_{1u}$(1), H$_g$(3), and H$_g$(4)  modes soften up
 to 14 GPa \cite{Snoke1992}.
Unusual softening of F$_{1u}$(1) and several Raman and infrared
inactive modes are also observed in other independent variable
pressure  measurements \cite{Huang1991, Aoki1991,
Martin1994,Meletov1995b}. Martin \emph{et al.} highlight
pressure-induced softening of  many inactive modes up to 2.5 GPa as
well  \cite{Martin1994}. Despite these experimental findings, no
comprehensive explanation for the unusual mode softening of
F$_{1u}$(1) and the aforementioned Raman active modes has been
advanced \cite{AokiNote}.

\begin{table}
   \caption{\label{tab_pressure}}
 Mode Gr\"{u}neisen parameters of the Raman, infrared
active, and inactive modes of C$_{60}$ in two different pressure
ranges: 0 - 0.4 and and augmented to include selected data from 0.5
- 2.5 GPa. The reported value of $\chi_T$ = 6.9 $\times$ 10$^{-2}$
GPa$^{-1}$ is used in the calculation \cite{Fischer1991}.
\\

   \begin{ruledtabular}
   \begin{tabular}{rcclrrlclcc}
   %1234567
 Mode     & $\omega$     &d$\omega$/dP     & Ref. & $\gamma_i$ %& $d\omega/dP$   &$\gamma_i$
 \\
 &(cm$^{-1}$) &   (cm$^{-1}$/GPa)        &    %& $(cm^{-1}/GPa)$&
          \\

 \hline
 \\H$_g$(1) & 272         & -15.00     &  \cite{Meletov1995a}    &   -0.80     %&1.1 &
 \\
H$_g$(2)  & 435         & -12.00    & \cite{Meletov1995a}    &   -0.40         %  & 2.4&
\\
A$_g$(1)  & 496         & -12.00      &\cite{Meletov1995a}    & -0.35        % & 0.94&
\\
F$_{1u}$(1)&526         & -1.60  & \cite{Huang1991}   & -0.04            % &-0.45 &
\\
F$_{1u}$(2)&576         & 2.40   & \cite{Huang1991}   &   0.06         %  & -&
  \\
H$_g$(3)   &710         &-12.00   &\cite{Meletov1995a}      &     -0.24         %&-0.55&
\\

H$_g$(5)   &1099        &-&\cite{Meletov1995a}\footnotemark[1]&-\\

F$_{1u}$(3)&1182        & 3.90  &\cite{Huang1991}     &    0.05          %&-      &
\\
H$_g$(6)   &1248        &-    &\cite{Meletov1995a}\footnotemark[1]          &   -           %&-      &
\\
F$_{1u}$(4)&1429        &6.10  &\cite{Huang1991}   &  0.06        %&-       &
\\

A$_g$(2)   &1467        &-18.00 &\cite{Meletov1995a} & -0.18\\
H$_g$(8)   & 1570       &-7.00  &\cite{Meletov1995a} &-0.06\\
&            &       &       & $\gamma_{av}\sim$ -0.19\footnotemark[2]  \\
&\\ \hline\\

H$_g$(4)   &772         & -2.70&\cite{Meletov1995b}\footnotemark[3]         &  -0.05\\
H$_g$(7)   &1422        &9.80   &\cite{Meletov1995b}\footnotemark[3] & 0.10     \\
H$_u$(3)& 668  & 0.62 & \cite{Martin1994}\footnotemark[3] & 0.01  \\
F$_{2u}$(2) &712 & -1.70 & \cite{Martin1994}\footnotemark[3] & -0.04\\
G$_{u}$(2)  &741 &-1.90 &  \cite{Martin1994}\footnotemark[3] & -0.04 &\\
F$_{2u}$(3)&796 & 0.73 & \cite{Martin1994}\footnotemark[3]& 0.01\\
H$_u$(4) & 824 & 3.00 & \cite{Martin1994}\footnotemark[3] & 0.05\\
G$_g$(2) &961 & 1.80 & \cite{Martin1994}\footnotemark[3]& 0.03\\

 &            &       &       & $\gamma_{av}\sim$-0.10\footnotemark[4]\\

 \footnotetext[1] {Ref. \cite{Comment5}}
\footnotetext[2]{Low pressure regime: 0 - 0.4 GPa.}
\footnotetext[3]{Slope between 0.4 - 2.4 GPa}
\footnotetext[4]{Inclusion of Raman and infrared active modes from
the low pressure regime as well as selected inactive modes with
pressure-dependent frequency shifts obtained from the high pressure
regime, as indicated in the Table. }

\end{tabular}
\end{ruledtabular}
\end{table}

Table~\ref{tab_pressure} summarizes our extended analysis of the
aforementioned pressure dependent vibrational studies of C$_{60}$,
undertaken as part of our search for microscopic evidence of
molecular level NTE in fullerenes. As discussed previously, many
modes soften with increasing pressure (especially between 0 - 0.4
GPa), giving negative Gr\"{u}neisen parameters for the majority of
modes.
 The size and sign of the Gr\"{u}neisen parameters are compatible with
 small but significant
expansion in intramolecular bond lengths as pressure is increased or
temperature is decreased \cite{Comment4}. Also, note that both
radial  and tangential modes exhibit pressure-induced softening
\cite{PressureNote}. The average value of the mode Gr\"{u}neisen
parameters  including most of optically active modes (upper portion,
Table~\ref{tab_pressure}) is  negative in this pressure range
($\gamma_{av}\sim$ -0.19). In order to further pin point the total
average mode Gr\"{u}neisen parameter for C$_{60}$, we extended this
analysis to include several optically inactive and two Raman active
modes over a somewhat larger pressure range  0.5 - 2.5 GPa
\cite{Martin1994}. The selected modes are shown in the lower portion
of Table~\ref{tab_pressure}. Thus, combining the data for these
modes in the higher pressure range (0.5 - 2.5 GPa) with the
aforementioned mode Gr\"{u}neisen parameter of the optically active
modes in the lower pressure range yields an averaged value of the
mode Gr\"{u}neisen parameter  that is also negative
($\gamma_{av}\sim$ -0.10).

Can we use this information to calculate the thermal expansion
coefficient, $\alpha$?
 Certainly, the  negative value of the  averaged Gr\"{u}neisen parameter
is directly related to a negative thermal expansion coefficient.
With precise knowledge of mode Gr\"{u}neisen parameters and phonon
densities of states for each mode, the calculation of thermal
expansion coefficient is straightforward (Eqn.~(1)).
 It is, however, important to point out that
isolated C$_{60}$ has 46 vibrational modes and that we are able to
evaluate Gr\"{u}neisen parameters for only a small subset  based
upon currently available data, despite the two different trials
presented in Table~\ref{tab_pressure}. To finesse this problem, we
designate $\alpha_{vib}$ as the vibrational part of the thermal
expansion coefficient for the optically-active modes in the low
pressure range (0 - 0.4 GPa) shown in Table~\ref{tab_pressure}.
Using the reported density of states \cite{Allen1999, Gompf1994},
each mode was weighted to calculate the total specific heat from
mode specific heat and finally our pseudo thermal expansion
coefficient, $\alpha_{vib}$ \cite{Ravindran2000}. Based upon this
data, we find $\alpha_{vib}$ to be on the order of -10$^{-5}$
K$^{-1}$. Inclusion of the inactive modes over the larger pressure
range (bottom portion of Table~\ref{tab_pressure}) does not modify
this general picture. Note that this value quantifies molecular
rather than conventional lattice effects.

Independent experimental support for this vibrational analysis and
the predictions of Kwon \emph{et al.} \cite{Kwon2004} comes from our
reanalysis of the extended x-ray absorption fine structure study of
Kr@C$_{60}$, performed at the Kr-edge of the encapsulated Kr atom at
77 and 300 K \cite{Ito2004}. X-ray absorption spectroscopy is a
sensitive and direct probe of molecular size and other local
effects. The inset of Fig.~\ref{fig_softening}(b) shows the
Kaiser-Bessel Fourier transformed data. A small shift in the peak
position (to higher values of R) is clearly evident at low
temperature and is consistent with the mode softening of various
infrared and Raman active modes. The Fourier analysis and fitting is
tricky, but the authors extract a Kr-C path distance of 3.540(7)
{\AA}
 at low temperature and
3.537(10) {\AA} at 300 K, indicating the possibility of a slightly
relaxed low temperature cage \cite{notesize}. Translating this
observation into a thermal expansion coefficient, we obtain on the
order of -10$^{-5}$ K$^{-1}$, similar to that extracted from our
vibrational analysis.

Combined, these experiments provide broad support for the proposed
molecular-level  NTE  in small carbon cage compounds.  In C$_{60}$
and Kr@C$_{60}$, selected modes soften as temperature is lowered
and/or  pressure is increased. The mode softening correlates
directly with the size and geometry of the fullerene cage  and NTE
at molecular level offers a universal explanation of both
temperature and pressure effects \cite{Kwon2004, PressureNote}. The
trend holds even in the presence of subtle guest-host interactions
in Kr@C$_{60}$. As discussed by Kwon \emph{et al.} \cite{Kwon2004},
low-temperature cage expansion may originate from an unusual
combination of structural and vibrational entropic effects. Hence,
NTE  may be an intrinsic property of fullerene and other small-cage
molecules, although in bulk measurements, the molecular-level effect
may be dominated by bulk lattice expansion. On the other hand,
molecular electronics applications exploit single molecular
properties. Understanding the thermal properties of functional,
assembled molecular-based materials presents many future
opportunities and challenges.

\section{Conclusion}

We report the the high-resolution far infrared vibrational
properties of C$_{60}$ and Kr@C$_{60}$. We observe that the
F$_{1u}$(1) mode softens with decreasing temperature in both
materials (300 - 10 K) and propose that the softening is correlated
 to low temperature expansion of fullerene cage. This finding is  consistent with
 previous temperature/pressure dependent vibrational spectroscopy and is also
 supported by our reanalysis of  temperature dependent  Kr extended x-ray
absorption fine structure data
\cite{Loosdrecht1992,Matus1993,Hamanaka1995,Horoyski1995,
 Meletov1995a,Chandrabhas1992,Snoke1992,Aoki1991,Martin1994,Meletov1995b,Huang1991,Ito2004}.
To further test this hypothesis, high pressure spectral results
\cite{Meletov1995a, Meletov1995b, Huang1991} were employed to
calculate the
 mode Gr\"{u}neisen parameters. Many of these  Gr\"{u}neisen parameters are
 negative. These results are consistent
with the recent prediction for molecular-level NTE  by Kwon {\it et
al.} \cite{Kwon2004}. Further work is needed to extend these ideas
to doped fullerenes and
 other curved molecular systems.

\section{Acknowledgements}

This project was supported by the US National Science Foundation
under Grant No. DMR-01394140 (UT) and No. CHE-0307168 (YU). We thank
D. Tom\'anek and Y.-K. Kwon for valuable discussions.


\begin{references}
\bibitem{F1u1}
Four infrared active modes are F$_{1u}$(1), F$_{1u}$(2), F$_{1u}$(3)
and F$_{1u}$(4) at 526, 576, 1182  and 1429 cm$^{-1}$ respectively.
\bibitem{spacegroup}
The low temperature space group is Pa$\bar{3}$.
\bibitem{Comment4}
All the modes under investigation are intramolecular modes and
hence, related to cage rather than lattice effects.
\bibitem{Comment5} The intermediate region could not be measured as
it has a strong Raman mode of diamond at
 1333.4 cm$^{-1}$ which is many orders of magnitude stronger
than the Raman modes of C$_{60}$.
\bibitem{PressureNote}
The pressure and temperature dependent behavior of lattice parameter
clearly shows that the reduction in the lattice constant at 5 K
($\sim$ 14.042 ${\AA}$) is already surpassed for pressure larger
than 0.41 GPa ($\sim$ 14.031 ${\AA}$) at room temperature
(Table~\ref{tab_temp}). Hence, we correlate the softening of modes
up to 0.4 GPa in pressure to the 5 K temperature regime .
\bibitem{notesize}
Note that the size of error bar is mainly due to fitting procedure.
Examination of the data (inset in Fig. 1(b)) clearly shows that the
peak of the radial distribution function is at a higher value of R
at low temperature (77 K).
\end{references}
\end{document}